\documentclass{pasj00}
\draft

\begin{document}
\SetRunningHead{Author(s) in page-head}{Running Head}
\Received{2000/12/31}
\Accepted{2001/01/01}

\title{Phase-reference VLBI Observations of the Compact Steep-Spectrum Source 3C 138 }


%
 \author{%
   Zhi-Qiang \textsc{Shen},\altaffilmark{1,2}
   L.-L. \textsc{Shang},\altaffilmark{1}
   D.-R. \textsc{jiang},\altaffilmark{1,2}
   H.-B. \textsc{Cai},\altaffilmark{1,3}
   and
   X. \textsc{Chen}\altaffilmark{1,3}}
 \altaffiltext{1}{Shanghai Astronomical Observatory, Chinese Academy of Sciences, Shanghai 200030, China}
 \altaffiltext{2}{Joint Institute for Galaxy and Cosmology (JOINGC) of SHAO and USTC, China}
 \altaffiltext{3}{Graduate School of Chinese Academy of Sciences, Beijing
100012, China}


\maketitle

\begin{abstract}

We investigate a phase-reference VLBI observation that was
conducted at 15.4~GHz by fast switching VLBA antennas between the
compact steep-spectrum radio source 3C 138 and the quasar PKS
0528+134 which are about 4$^\circ$ away on the sky. By comparing
the phase-reference mapping with the conventional hybrid mapping,
we demonstrate the feasibility of high precision astrometric
measurements for sources separated by 4$^\circ$. VLBI
phase-reference mapping preserves the relative phase information,
and thus provides an accurate relative position between 3C 138 and
PKS 0528+134 of $\Delta\alpha=-9^m46^s.531000\pm0^s.000003$ and
$\Delta\delta=3^\circ6^\prime26^{\prime\prime}.90311\pm0^{\prime\prime}.00007$
(J2000.0) in right ascension and declination, respectively. This
gives an improved position of the nucleus (component A) of 3C 138
in J2000.0 to be RA=$05^h 21^m 9^s.885748$ and Dec=$16^\circ 38'
22''.05261$ under the assumption that the position of calibrator
PKS 0528+134 is correct. We further made a hybrid map by
performing several iterations of CLEAN and self-calibration on the
phase-referenced data with the phase-reference map as an input
model for the first phase self-calibration. Compared with the
hybrid map from the limited visibility data directly obtained from
fringe fitting 3C 138 data, this map has a similar dynamic range,
but a higher angular resolution. Therefore, phase-reference
technique is not only a means of phase connection, but also a
means of increasing phase coherence time allowing self-calibration
technique to be applied to much weaker sources.

\end{abstract}

\section{Introduction}

The self-calibration algorithm for the interferometer visibility
has proved to work well in the high signal-to-noise ratio (SNR)
case. It uses the closure-phase relations to calibrate visibility
phase, and then the absolute positional information is lost after
the phase self-calibration. Furthermore, when the SNR is low, the
visibility phases will have a large scatter. No signal can be
detected if the SNR is below a flux limit imposed by the
relatively short coherence time. Therefore, to study the structure
of very weak radio sources we should take advantage of
phase-reference mapping (Alef 1989). With the publication of the
first phase-reference mapping from in-beam observations (Marcaide
et al 1984) and from switched observations (Alef 1988), phase
referencing technique has gradually become a powerful tool for
imaging weak radio sources (Beasley \& Conway 1995). This
technique can hopefully work without additional atmospheric or
ionospheric calibration measurements, because these effects can be
well modelled from the observation of nearby, strong reference
sources. The applicability of this technique relies on the fact
that most of the instrumental and propagation effects are common
for both reference and target sources which are sufficiently close
on the sky. While hybrid mapping in VLBI provides information only
about the relative position between different features in a given
source, the phase-reference observations can provide precise
position of the target source with respect to the external
reference at sub-milliarcsecond accuracy (Ros 2004).

In this paper, we present a phase-reference VLBI study of the
compact steep spectrum (CSS) source 3C 138. The target source 3C
138 (=4C 16.12=J0521+166), with a priori position of RA $05^h 21^m
9^s.88603$ and Dec $16^\circ 38' 22''.0519$ (J2000.0), is
identified as a quasar with $m_{v}=18.84$ and $z=0.759$ (Hewitt \&
Burbidge 1989). The reference source was PKS 0528+134 which is a
compact $\gamma$-ray-bright quasar, located at RA $05^h 30^m
56^s.41675$, Dec $13^\circ 31' 55''.1495$ (J2000.0) and has a
redshift $z=2.06$. The angular separation of 3C 138 from PKS
0528+134 on the sky is about $3.9^\circ$. In Sect. 2 we first
describe the observations, and then present two new 15.4 GHz VLBI
maps of 3C 138 obtained from the phase-reference related mapping.
In Sect. 3 we compare and discuss maps made from three different
methods, namely hybrid mapping, phase-reference mapping and a
combination of phase-reference with CLEAN and self-calibration.
From the phase-reference map, we also estimate the relative
separation between the target source 3C 138 and the calibrator PKS
0528+134, from which an improved coordinate of 3C 138 is obtained.
General conclusions are presented in Sect. 4.

\section{Observations and Data Reduction}

We carried out quasi-simultaneous VLBI observations of 3C 138 on
2001 August 20, using ten 25m VLBA antennas. The phase referencing
observations at 15.4 GHz were interleaved with observations at
three other frequencies (2.3, 5 and 8.4 GHz) at which no
phase-referencing mode was adopted. The main purpose of the four
quasi-simultaneous multi-frequency VLBI observations is to
identify the center of activity of the CSS superluminal source 3C
138 (Shen et al. 2005). The phase-reference VLBA observation at
15.4 GHz was performed through a rapid switching between the
target source 3C 138 and the reference source PKS 0528+134. The
switching cycle time is 100s, consisting of 32s on PKS 0528+134,
8s for slewing antenna, 52s for 3C 138, and another 8s for slewing
back antenna. Correlation was done at the VLBA correlator in
Socorro, New Mexico (USA).

The post-correlation data reduction was performed in the NRAO AIPS
and the Caltech Difmap packages. A priori visibility amplitude
calibration was first made using system temperatures and gain
curves from each antenna. We then picked up one high SNR scan of
PKS 0528+134 to calibrate the instrumental delays and phase
offsets between IF channels. VLBA-LA (Los Alamos at New Mexico)
was used throughout as the reference station. Then we did
fringe-fitting (FRING in AIPS) on the calibrator PKS 0528+134's
visibility data and mapped it with a peak-to-rms dynamic range of
$\sim$1500 (see Fig. 1 of Cai et al. 2006).

Similarly, we also did a global fringe fitting to 3C 138 data
directly in AIPS, and then made a hybrid map (see Fig. 2 of Shen
et al. 2005) from the detected visibility data. Due to the
weakness of the radio emission, we only detected fringes on the
baselines to six of ten VLBA stations (FD, KP, LA, NL, OV, PT). We
then averaged all channels within each IF and exported the data to
Difmap. In Difmap, we first averaged the visibility data to a
20-second grid with the weights calculated from the scatter in the
data. The averaged data were then phase self-calibrated with a
point source model, followed by several iterations of CLEAN and
phase-only self-calibration. We began with uniform weighting, and
later switched to the natural weighting in order to reveal more
extended structure. For amplitude calibration, we only made an
overall constant gain correction to each station.

To investigate VLBI phase-reference observation, we further made
two other maps using the phase-referenced data as described below.

\subsection{Phase-reference Mapping}

First it should be noted that PKS 0528+134 at 15.4~GHz is not an
ideal point source (see Fig.1 of Cai et al. 2006). There is
extended emission towards the north-east, which accounts for about
11\% of the total integrated flux density at 15.4~GHz. In order to
remove such structural phase effects from the reference source PKS
0528+134, all clean components in the hybrid map of PKS 0235+164
were fed back into the phase self-calibration process as an input
model to refine the estimate of the antenna-based residuals. By
applying to the 3C 138 data these resultant phase-like solutions
determined from fringe-fitting the PKS 0528+134 data (after
minimizing the structural effect), we obtained a set of
phase-referenced data, which were then transferred into Difmap.
These phase-referenced visibility data were time averaged, Fourier
inverted and deconvolved (without any phase self-calibration) to
produce the first phase-reference map of 3C 138 at 15.4 GHz (Fig.1
(a)).

\subsection{Hybrid Mapping of the Phase-referenced Data}
This time, after obtaining the phase-referenced visibility data,
we further carried out phase self-calibration using the
phase-reference map Fig.1 (a) rather than the point source model.
Then we simply follow the standard self-calibration algorithm to
produce a high-resolution 15.4 GHz VLBI map of 3C~138 (Fig.1 (b)).

\section{Discussion}

To yield a quantitative description of source structure exhibited
in these maps, we fitted the visibility data with circular
Gaussian components. The results are listed in Tables 1 and 2
along with some parameters of each map in Table 3.

\subsection{Comparison of Three Maps}

As the consequence of the coherence loss due to the temporal phase
fluctuation and large angular separation between the target and
reference sources, the phase-referenced map (Fig.1(a)) reveals
only two brighter components (A and B1) of the four well detected
VLBI components (A, B1, B2, and C) shown in the hybrid maps of
3C~138 by Shen et al. (2005). Even though, this clearly
demonstrates the feasibility of high precision astrometric
measurements for sources separated by 4$^\circ$. The central four
VLBI components have also been consistently seen at 5.0 and 8.4
GHz in 2001 August (Shen et al. 2005), in good agreement with the
5.0 GHz linear polarization images in three epochs from 1998
September to 2002 October (Cotton et al. 2003).

Direct global fringe-fitting on 3C 138 threw away a lot of
visibility data that are below the SNR threshold, and thus are not
used in the hybrid mapping. It should be noted that in the case of
VLBI observation of 3C~138, most of these discarded data are on
longer baselines because of the resolution effect. This results in
a smaller uv-coverage (as shown in the left plot of Fig. 2) and
thus a poorer spatial resolution, i.e., a larger beam (see Table
3). In contrast, the phase-referencing method would keep all the
observational visibility data of the target source, i.e., preserve
the uv-coverage. So the beam size of the phase-reference map
Fig.1(a) stands for the full resolution of the whole data sets
(shown in the right plot of Fig.2). And since the phase
self-calibration process (SELFCAL) in Difmap can keep the
uv-coverage even if the SNR is low, so the beam size in Fig.1(b)
is essentially the same as that in Fig.1(a) (Table 3). The
uv-coverage plots in Fig.2 (also refer to Table 3) reveal that the
resolutions of Fig.1(a) and Fig.1(b) are about 4 and 2 times
higher than that of Fig.2 of Shen et al. (2005) in right ascension
and declination, respectively.

Moreover, because several iterations of phase self-calibration
have been performed in making Fig. 1(b) from the phase-referenced
visibility data, Fig.1(b) achieves a similar dynamic range as the
hybrid map (Fig. 2 of Shen et al. 2005). It also clearly displays
three compact components (A, B1, and B2) with a diffuse emission
at the location of component C which is heavily resolved at a full
resolution here. For a consistent check, we also made a hybrid map
using the truncated phase-referenced data which have the same
uv-coverage as that of Fig.2 of Shen et al. (2005), and
eventually, a qualitatively similar image was obtained.

As can be seen in Table 3, both the peak flux density and dynamic
range are the lowest in the phase-reference map (Fig.1(a)). This
implies a coherence loss of about 84\%, mainly due to the temporal
phase fluctuation within each scan interval of 52~s for 3C~138.
The coherence of an interferometer is a measurement of the phase
stability of the entire system (Rogers and Moran 1981), it is the
ratio of the time-integrated fringe amplitude to the instantaneous
fringe amplitude (Linfield et al. 1989). Coherence losses anywhere
in the system will degrade the SNR and cause a lower peak flux
density. In addition, the large ($4^\circ$ or so) angular
separation of the target and reference sources will introduce some
uncertainties in referencing the phases since the radiation from
both sources may propagate through a slightly different medium.
Furthermore, the inaccurate angular separation model in the
correlator also limits the dynamic range of the phase-reference
map.

\subsection{Relative Position between 3C~138 and PKS~0528+134}

There is an offset of the map center (0,0) from either of the
detected compact VLBI components A and B1 in the phase-reference
map Fig.1(a). This kind of offset has been also seen in many other
phase-reference observations (Rioja \& Porcas 2000; Alef 1988).
This is mainly due to the deviation of the true separation between
the target and the reference from the separation model adopted in
the correlator. If the separation between the target and
calibrator sources is exactly the same as the model used in the
correlator, the offset would be zero. Therefore, by determining
such kind of positional deviation from the phase-reference
observation, we can obtain a more accurate correction to the
relative position known at the correlator. It should be mentioned
that although a similar offset is also seen in Fig.1(b), it cannot
be used to estimate the relative position since some iterations of
phase self-calibration have been invoked in the mapping process.
The absolute positional information is not kept in the hybrid
mapping (Fig. 2 in Shen et al. 2005) either, as the phase
self-calibration has already done. The phase-reference map is the
unique one that can be used to derive an accurate measurement of
the absolute positional information.

Since both the reference source PKS 0528+134 and target source
3C~138 are non-point sources, it is necessary to determine the
suitable reference points for the differential astrometric
measurement of the separation between them. Based on the
high-resolution spectra as well as the variability study,
component A has been identified as the location of the central
activity in 3C 138 (Shen et al. 2001; 2005), and thus will be used
as a position reference point for 3C~138 in the following
astrometric analysis. For the core-dominated blazar PKS 0528+134
(Cai et al. 2006), the peak of the compact core at the phase
center is chosen as the reference point. Ideally, a reference
point should correspond to the peak of a strong, unresolved
component, which is well separated from other radio emission
within the source structure (Rioja \& Porcas 2000). As a result,
the positional corrections to the correlator model of 3C~138 can
be estimated from the position of reference feature (component A)
in the phase-reference map (Fig. 1(a)), which are of
$-4.05\pm0.05$ mas and $0.71\pm0.07$ mas in right ascension and
declination, respectively. Transforming the offset to the J2000.0
coordinates, we have
$\delta(\Delta\alpha)=-0^s.000280\pm0^s.000003$ and
$\delta(\Delta\delta)=0^{\prime\prime}.00071\pm0^{\prime\prime}.00007$,
respectively. Here, to convert the correction to right ascension
in second, we have divided the correction to right ascension in
mas by the factor 15cos$\delta_m$, where $\delta_m$ is the mean
declination of 3C 138 and PKS 0528+134 ($\sim15^\circ$).

The quoted standard errors are mainly associated with the
uncertainties in determining the position of the peak in component
A using the AIPS task JMFIT. For comparison, the errors due solely
to the finite SNR of component A in 3C 138 are estimated to be
0.02 and 0.03 mas in right ascension and declination,
respectively, from the expression
$\sigma=(1/2\pi)\times(1/SNR)\times(\lambda/D)$ in radians
(Thompson, Moran \& Swenson 1986) by choosing the equivalent
baseline length D of 2$\times10^8\lambda$ and
1.5$\times10^8\lambda$ along E-W and N-S directions (see Fig. 3),
respectively.

Taking into account the separation in the a priori coordinates of
the correlator model ($-9^m 46^s.53072$ (RA), $3^\circ 6'
26''.9024$ (Dec)), we can obtain a revised separation of 3C 138
from PKS 0528+134 in J2000.0:
$\Delta\alpha=-9^m46^s.531000\pm0^s.000003$ and
$\Delta\delta=3^\circ6^\prime26^{\prime\prime}.90311\pm0^{\prime\prime}.00007$.
By assuming there is no error in the a priori position of quasar
PKS 0528+134, we can further derive an improved coordinate of the
nucleus (component A) of 3C 138 to be:

RA(J2000.0)=$05^h 21^m 9^s.885748$

Dec(J2000.0)=$16^\circ 38' 22''.05261$ .

However, it should be noted that errors caused by the incomplete
known tropospheric delays, coupled with a large angular separation
between the target and calibrator at lower elevations, could be a
few times larger than those due to the identification of the
centroid on the map. Such systematic errors can be greatly reduced
by choosing a calibrator much closer to the target, or by using
more than one calibrator in the vicinity of the target, with a
weighting depending on their position with respect to the tagret,
to determine the angular dependence of the phase (Fomalont 2005).
The limitation of the single calibrator technique for such a not
so closely spaced target-calibrator pair is quite apparent, as
evident by a large phase decorrelation (see Sect. 3.1).

\section{Conclusions}

We report on a phase-reference VLBI study of the compact steep
spectrum source 3C 138 at 15.4~GHz. The observation was conducted
using the phase-referencing technique by fast switching between 3C
138 and the bright compact quasar PKS 0528+134.

The first phase-referenced map of 3C~138 was successfully made at
15.4 GHz, which demonstrates the feasibility of high precision
astrometric measurements for sources separated by 4$^\circ$. This
phase-reference VLBI map retains the relative phase information,
and we have updated the relative position in J2000.0 between 3C
138 and PKS 0528+134 to be
$\Delta\alpha=-9^m46^s.531000\pm0^s.000003$ and
$\Delta\delta=3^\circ6^\prime26^{\prime\prime}.90311\pm0^{\prime\prime}.00007$,
respectively. This provides an improved position of 3C 138 to be
$05^h 21^m 9^s.885748$ and $16^\circ 38' 22''.05261$ (J2000.0) in
right ascension and declination, respectively, under the
assumption that there is no error in the position of the strong
reference source PKS 0528+134.

By comparing the hybrid maps made from the visibility data with
and without applying phase-reference technique, we found that both
maps have a similar dynamic range, but the hybrid map from the
phase-referenced data has a higher angular resolution since more
visibility data are recovered. Therefore, the phase-reference
technique is not only a means of phase connection, but also a
means of increasing phase coherence time to improve the detection
to much weaker sources.

\vspace{1cm} We thank an anonymous referee for helpful comments on
the earlier manuscript. This work is supported in part by the
National Natural Science Foundation of China under grant 10573029,
and sponsored by Program of Shanghai Subject Chief Scientist
(06XD14024). Z.-Q. Shen acknowledges the support by the
One-Hundred-Talent Program of Chinese Academy of Sciences.


\clearpage
\begin{table}
  \caption{Flux density in (mJy) of each component from Circular
Gaussian model fit}\label{tab:first}
  \begin{center}
\begin{tabular}{clclclclclcl}
\hline\noalign{\smallskip}
Map ID & &A & &B1 & &B2 & &C   \\
\hline\noalign{\smallskip}
Fig.1 (a) & &6.84 & &49.20 & &$\cdot\cdot\cdot$ & &$\cdot\cdot\cdot$   \\
Fig.1 (b) & &30.88     & &65.14   & &21.36  & &$\cdot\cdot\cdot$ \\
Fig.2 in Shen et al. (2005) & &26.50  & &66.86  & &24.10 & &11.26
\\\noalign{\smallskip}\hline
\end{tabular}
  \end{center}
\end{table}
\begin{table}
  \caption{Relative position (r, P.A.) in (mas, deg) of each
component with respect to component A at (0,0)}\label{tab:second}
  \begin{center}
\begin{tabular}{clclclclclcl}
\hline\noalign{\smallskip}
Map ID  & &B1 & &B2 & &C  \\
        & &(mas, $^\circ$) & &(mas, $^\circ$) & & (mas,$^\circ$) \\
\hline\noalign{\smallskip}
Fig.1 (a) & &6.28, 89.0 & &$\cdot\cdot\cdot$ & &$\cdot\cdot\cdot$ \\
Fig.1 (b) & &6.33, 88.8 & &4.88, 94.2 & &$\cdot\cdot\cdot$ \\
Fig.2 in Shen et al. (2005) & &6.36, 89.9 & &4.93, 94.3 &&9.51,
99.6
\\\noalign{\smallskip}\hline
\end{tabular}
  \end{center}
\end{table}
\begin{table}
  \caption{Parameters measured in three maps}\label{tab:third}
  \begin{center}
\begin{tabular}{clclclclclclcl}
\hline\noalign{\smallskip}
Map ID  &&restoring beam & &rms noise & &peak & &dynamic \\
&&(mas $\times$ mas, deg) & &(mJy/beam) & &(mJy/beam) & &range \\
\hline\noalign{\smallskip}
Fig.1 (a) &&$0.834\times0.431$, $-$9.67 & &0.736 & &5.7 & &7.7 \\
Fig.1 (b) &&$0.835\times0.430$, $-$9.91 & &0.439 & &35.1 & &80.0 \\
Fig.2 in Shen et al. (2005) & &$1.77\times1.38$, $-$20.9 & &0.631
& &62.5 & &99.0
\\\noalign{\smallskip}\hline
\end{tabular}
  \end{center}
\end{table}

  \clearpage
  \begin{figure}
\centering \vspace{430pt} \includegraphics{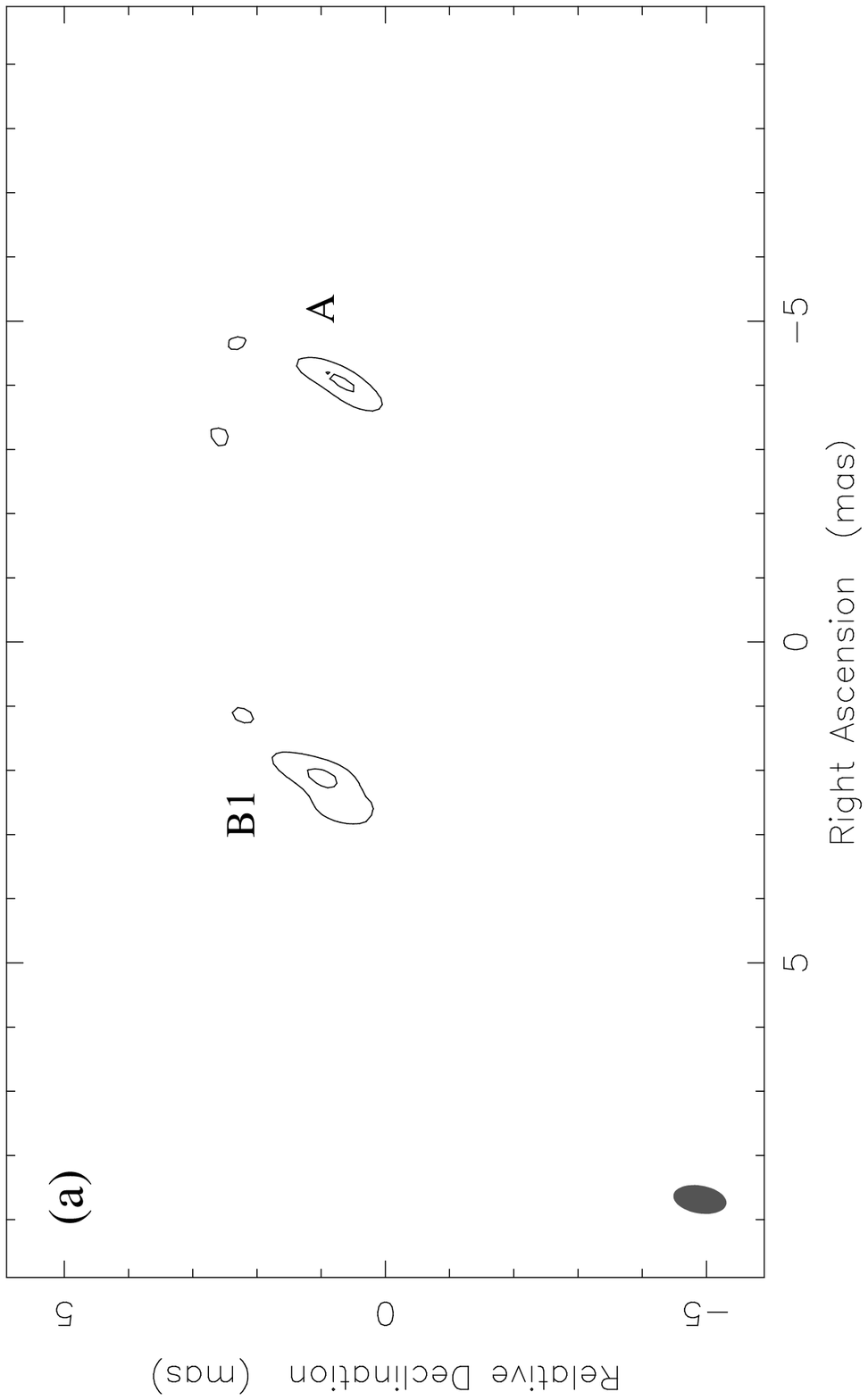}
\includegraphics{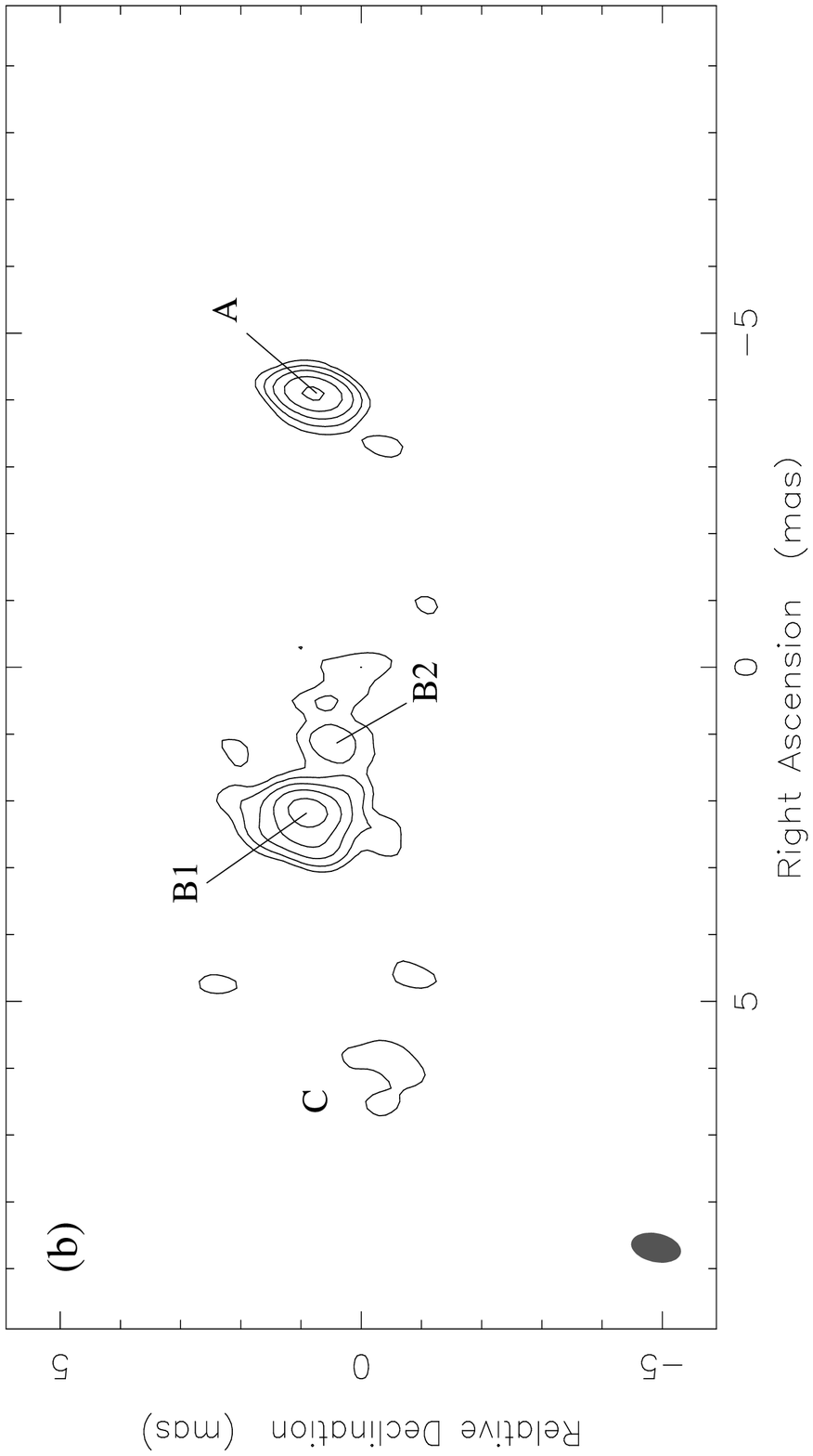} \caption{Two 15.4 GHz VLBI maps of 3C~138
made from the phase-referenced visibility data using the
phase-reference mapping ((a): contours: 2.5 mJy/beam$\times(1,
2)$, rms noise: 0.736 mJy/beam) and, the combination of
phase-reference mapping with CLEAN and self-calibration ((b):
contours: 1.5 mJy/beam $\times (1, 2, 4, 8, 16)$, rms noise: 0.439
mJy/beam) \label{Fig:figure1} }
\end{figure}
  \clearpage
  \begin{figure}
\centering \vspace{185pt} \includegraphics{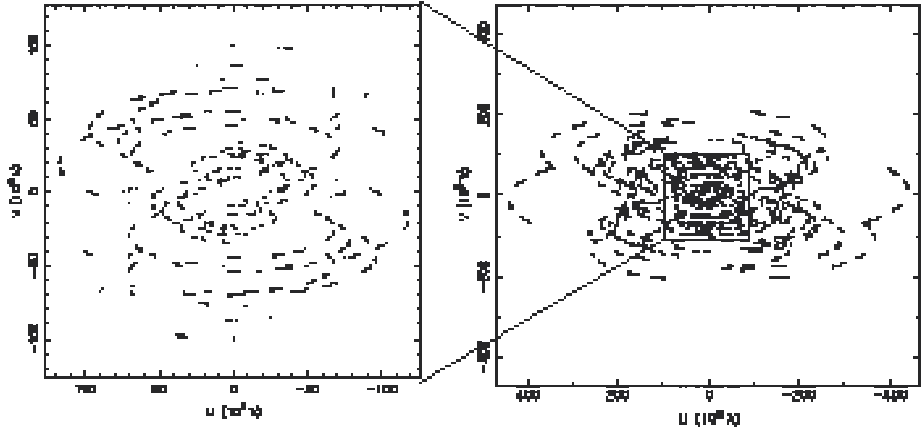} \caption{The
uv-coverage of the visibility data used in the hybrid mapping
(left panel) and the phase-reference mapping (right panel)
\label{Fig:figure2} }
\end{figure}
\end{document}